\documentclass[aps,prl,reprint,superscriptaddress,
]{revtex4-2}
\usepackage{graphicx}
\usepackage{dcolumn}
\usepackage{bm}
\usepackage{colortbl}
\usepackage{float}
\usepackage[table,dvipsnames]{xcolor}
\usepackage{makecell}
\usepackage[mathscr]{euscript}
\usepackage{ulem}
\usepackage{multirow}
\usepackage{braket}
\usepackage[pagewise]{lineno}
\usepackage{enumitem}
\usepackage{amssymb}
\usepackage{makecell}
\usepackage{tikz}
\usepackage{bbding}
\usepackage{lineno}

\begin{document}

\title{Isotropic Superconductivity in Room-temperature Superconductor LaSc$_{2}$H$_{24}$}

\author{Zefang Wang}
\affiliation{Key Laboratory of Material Simulation Methods $\&$ Software of Ministry of Education, College of Physics, Jilin University, Changchun, 130012, China}

\author{Wenbo Zhao}
\affiliation{Key Laboratory of Material Simulation Methods $\&$ Software of Ministry of Education, College of Physics, Jilin University, Changchun, 130012, China}

\author{Yuan Ma}
\affiliation{Key Laboratory of Material Simulation Methods $\&$ Software of Ministry of Education, College of Physics, Jilin University, Changchun, 130012, China}

\author{Hanyu Liu}
\email{hanyuliu@jlu.edu.cn}
\affiliation{Key Laboratory of Material Simulation Methods $\&$ Software of Ministry of Education, College of Physics, Jilin University, Changchun, 130012, China}
\affiliation{State Key Laboratory of High Pressure and Superhard Materials, College of Physics, Jilin University, Changchun, 130012, China}
\affiliation{International Center of Future Science, Jilin University, Changchun 130012, China}

\author{Yanming Ma}
\email{mym@jlu.edu.cn}
\affiliation{School of Physics, Zhejiang University, Hangzhou 310027, China}
\affiliation{Key Laboratory of Material Simulation Methods $\&$ Software of Ministry of Education, College of Physics, Jilin University, Changchun, 130012, China}
\affiliation{State Key Laboratory of High Pressure and Superhard Materials, College of Physics, Jilin University, Changchun, 130012, China}
\affiliation{International Center of Future Science, Jilin University, Changchun 130012, China}
\date{\today} 

\begin{abstract}
The discovery of LaSc$_{2}$H$_{24}$ represents a milestone in the quest for room-temperature superconductivity, yet the microscopic mechanism underlying its superior performance remains unclear. Through a comprehensive revisit of theoretical calculations, we uncover a pivotal transition from the anisotropic two-gap superconductivity of LaH$_{10}$ to the isotropic single-gap superconductivity in LaSc$_{2}$H$_{24}$ upon the introduction of scandium, thereby enhancing the superconducting critical temperature ($T_\mathrm{c}$). This enhancement is rooted in a critical dual role of Sc $3d$ electrons: i) the Sc-derived Jahn-Teller effect promotes hydrogen metallization via the elongation of specific interlayer H-H bonds and enhances electron-phonon coupling (EPC) through the softening of associated phonon modes; ii) Sc $3d$ electrons reconstruct the electronic structure into an MgB$_{2}$-like configuration, generating novel Sc-H-Sc $\sigma$- and $\pi$-bonding states with EPC strengths comparable to LaH$_{10}$. Crucially, the pronounced hybridization between Sc and the hydrogen cages effectively unifies these two contributions on the Fermi surface. This Sc-induced gap unification bridges the high-EPC H-H states with widespread Sc-H states, establishing an isotropic single-gap nature with a large overall EPC strength. Our findings identify this Sc-induced gap unification as the fundamental mechanism for achieving room-temperature superconductivity in LaSc$_{2}$H$_{24}$, offering a theoretical blueprint for the future design of superior superconducting hydrides.
\end{abstract}

\maketitle

The pursuit of room-temperature superconductivity in clathrate metal hydrides originated with the prediction of CaH$_{6}$ in 2012 \cite{CaH6_2012}, a structure synthesized a decade later by two independent teams \cite{CaH6_2022_Ma, CaH6_2022_Jin}. Subsequent efforts shifted towards rare-earth metal hydrides, leading to the prediction of the near-room superconductor LaH$_{10}$ \cite{LaH10_2017_Peng, LaH10_2017_Liu}, which was confirmed with the superconducting critical temperature ($T_{\mathrm{c}}$) up to 260 K at 188 GPa by two experimental groups \cite{LaH10_2019_Eremets, LaH10_2019_Hemley}. The realization of LaH$_{10}$ not only lifted the record of $T_{\mathrm{c}}$, but also established clathrate metal hydrides as the most promising candidates for room-temperature superconductivity. Inspired by these breakthroughs, numerous clathrate hydrides have been theoretically predicted and successfully synthesized \cite{Review_2023_Sun, Review_2020_Lv, Review_2020_Eremets, Review_2021_Yang, Review_2022_Zhong, Review_2022_Cui}. Until recently, the theoretically predicted room-temperature superconductor LaSc$_{2}$H$_{24}$ was realized in the experiment \cite{LaSc2H24_2024_He, RTSC}.

LaSc$_{2}$H$_{24}$ is designed based on the strategy of tuning superconductivity through doping metals in hydrogen-rich superconductors, a concept first proposed in the theoretical design of $Fd\overline{3}m$ Li$_{2}$MgH$_{16}$ with a $T_{\mathrm{c}}$ up to 473 K \cite{Li2MgH16_2019_Sun}. Doping Sc into the La-H system results in the formation of LaSc$_{2}$H$_{24}$, which exhibits thermodynamic stability and superconductivity exceeding LaH$_{10}$. This room-temperature superconductor crystallizes in the $P6/mmm$ space group, featuring a lanthanum-scandium sublattice that resembles the MgB$_{2}$ lattice. Although subsequent theoretical research has been conducted \cite{AB2H24_Search_Jiang, AB2H24_Jiang, AB2H24_PCCP}, the microscopic mechanism of such room-temperature superconductivity remains elusive, especially the $T_{\mathrm{c}}$ enhancement induced by the introduction of scandium.

Herein, we present a comprehensive first-principles study of LaSc$_{2}$H$_{24}$ at 250 GPa to elucidate the origin of its record-breaking superconductivity. Our analysis of crystal orbitals and electronic structure indicates that the anisotropic hybridization between Sc $3d$ electrons and hydrogen cages induces a pronounced crystal-field effect, triggering an apparent elongation of interlayer H-H bonds ($\sim 1.2$ {\AA}) derived by Jahn Teller effect. This structural modulation enriches the H-H anti-bonding states at the Fermi level ($\varepsilon_{\mathrm{F}}$). Simulations of phonon properties and electron-phonon coupling (EPC) reveal significant phonon softening at the $q_\mathbf{K}$ wave vector. These softened modes, arising from the instability of elongated H-H bonds and the absence of Sc-H hybridization, substantially enhance the EPC contribution of the hydrogen atoms. Furthermore, the introduction of scandium reconstructs the electronic structure into a configuration analogous to MgB$_{2}$, characterized by Sc-H-Sc $\sigma$- and $\pi$-bonding states at $\varepsilon_{\mathrm{F}}$. These Sc-H states exhibit EPC strengths comparable to those of widespread La-H states in LaH$_{10}$. Crucially, by solving the equations of the anisotropic Migdal-Eliashberg equation \cite{Eliashberg, EPW1, EPW2, EPW3, EPW4}, we demonstrate that the introduction of scandium drives a pivotal transition from the anisotropic two-gap superconductivity of LaH$_{10}$ to an isotropic single-gap superconductivity in LaSc$_{2}$H$_{24}$. This single-gap characteristic implies the effective merging of Sc-H and H-H states on the Fermi surface (FS), a unification rooted in the pronounced anisotropic hybridization between Sc $3d$ electrons and the hydrogen cages. By bridging the high-EPC H-H states with the widespread Sc-H states, this merging prevents the separation of the superconducting gap ($\Delta_{n\mathbf{k}}$) and establishes a continuous distribution of large EPC constants ($\lambda_{n\mathbf{k}}$). Our findings identify this Sc-induced gap unification as the fundamental mechanism for achieving room-temperature superconductivity in LaSc$_{2}$H$_{24}$, providing a strategic blueprint for the design of superior ternary hydrides.


The structure of LaSc$_{2}$H$_{24}$ is characterized by fully enclosed La@H$_{30}$ cages and partially enclosed Sc@H$_{24}$ cages \cite{Review_2023_Sun}. Notably, metal atoms adopt a MgB$_{2}$-type lattice, with La and Sc atoms occupying the Mg and B Wyckoff positions, respectively. This $P6/mmm$ high-symmetry sublattice dictates the shape of the surrounding hydrogen framework, giving rise to remarkable structural anisotropy. The hydrogen framework is classified into three distinct layers parallel to the (001) crystal plane: H$_{\mathrm{I}}$ atoms form regular hexagons ($d_\mathrm{H-H}$ = 1.11 {\AA}) positioned directly above the La sites; H$_{\mathrm{III}}$ atoms, carrying the maximum negative charge (-0.16 $e^{-}$/atom), form triangles ($d_\mathrm{side}$ = 1.34 {\AA}) located directly above the Sc sites; H$_{\mathrm{II}}$ atoms are arranged in short pairs ($d_\mathrm{H-H}$ = 1.09 {\AA}), situated between the H$_{\mathrm{I}}$ and H$_{\mathrm{III}}$ layers [Fig. 1]. The formation of this anisotropic framework is rooted in the interaction between Sc atoms and the hydrogen cages. Maximally localized Wannier functions (MLWFs) \cite{MLWF1, MLWF2} [Fig. S1] clearly demonstrate significant orbital overlap between the Sc $3d$ electrons and the surrounding H$_{24}$ cages. Further analysis of the orbital-projected band structure [Fig. S2] indicates strong low-lying hybridization between the hydrogen atoms and specific Sc $3d$ orbital components ($3d_\mathrm{zx}$, $3d_\mathrm{zy}$, $3d_\mathrm{xy}$, and $3d_\mathrm{x^{2}-y^{2}}$) at energies of approximately $(\varepsilon_{\text{F}}-8)$ eV. As this evident crystal-field splitting lowers the energy of the crystal, Sc $3d$ electrons dictate the structural anisotropy of the hydrogen framework and contribute to the thermodynamic stability of $P6/mmm$ LaSc$_{2}$H$_{24}$ among La-Sc-H ternary hydrides \cite{LaSc2H24_2024_He}.

\begin{figure}[htb]
    \centering
    \includegraphics[width=1.0\linewidth]{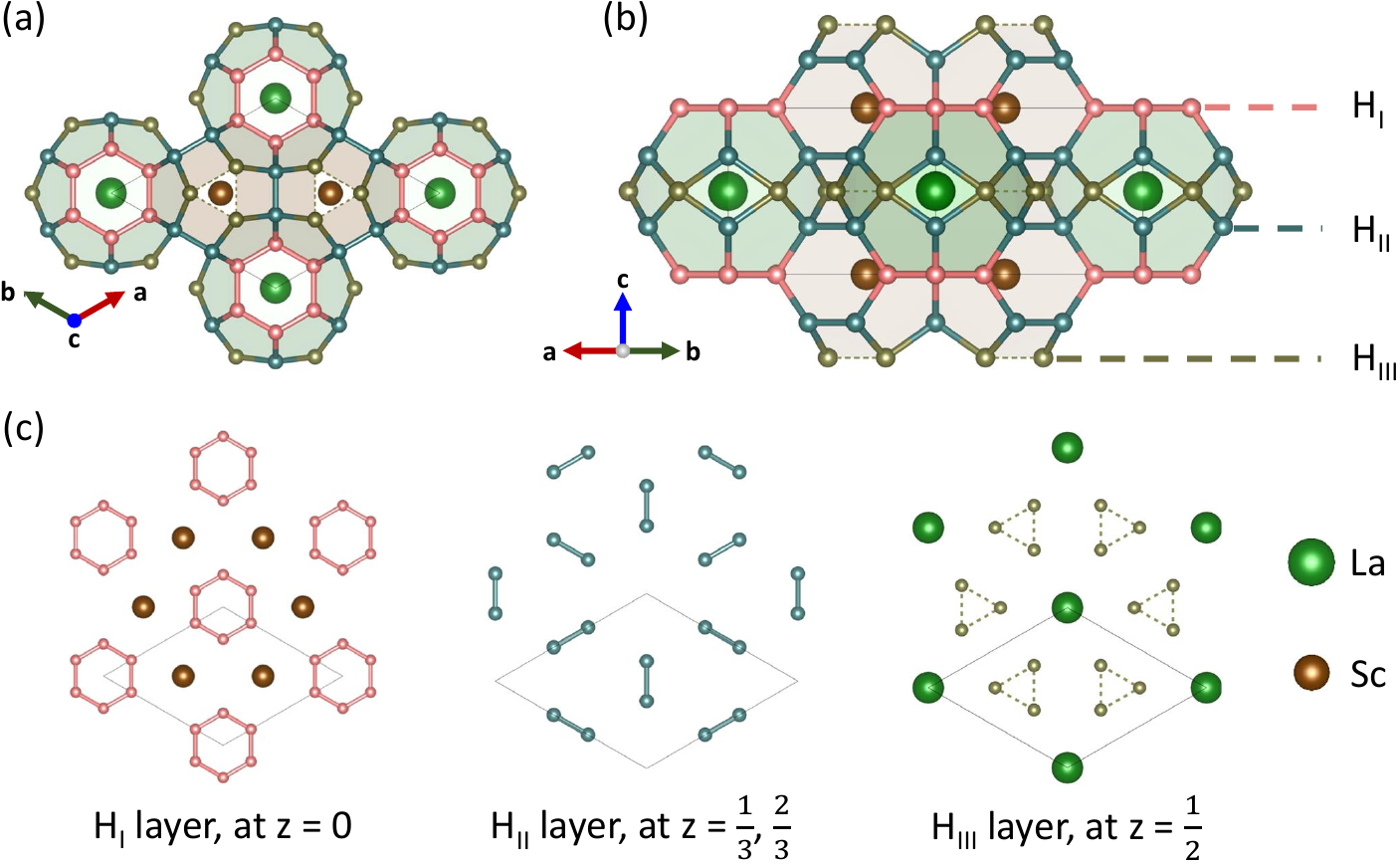}
    \caption{Optimized structure of the $P6/mmm$ LaSc$_{2}$H$_{24}$ crystal, shown in the (a) top and (b) side view. (c) The atom layers along the (001) plane at various $z$ heights ($z$ = 0, $\frac{1}{3}$, $\frac{1}{2}$, $\frac{2}{3}$). Hydrogen atoms are categorized into three inequivalent Wyckoff positions (H$_{\mathrm{I}}$, H$_{\mathrm{II}}$, H$_{\mathrm{III}}$).}
    \label{fig:Fig. 1}
\end{figure}

The significant disparity in ionic radii, coupled with the Jahn-Teller effect derived from Sc $3d$ electrons, directly modulates the structure of the hydrogen framework. Specifically, the anisotropic structural deformation increases the separation between hydrogen layers, causing a pronounced elongation in the interlayer H$_{\mathrm{I}}$-H$_{\mathrm{II}}$ and H$_{\mathrm{II}}$-H$_{\mathrm{III}}$ bonds to approximately 1.20 {\AA}. Examinations employing the electron localization function (ELF) \cite{ELF} and crystal orbital Hamiltonian population (COHP) \cite{COHP} reveal that these elongated bonds are pivotal for enhancing hydrogen metallization. As illustrated in Table S1 and Table S2, the ELF minimum values along these interlayer bonds range from 0.5 to 0.6, approaching the characteristic of a free electron gas. These metallic interactions ensure that H-H anti-bonding states dominate the Fermi level ($\varepsilon_{\mathrm{F}}$). COHP analysis demonstrates that H$_{\mathrm{I}}$-H$_{\mathrm{II}}$ anti-bonding states contribute prominently to the density of states (DOS) at $\varepsilon_{\mathrm{F}}$ [Fig. S3]. For comparison, LaH$_{10}$ also exhibits metallic H-H interactions with a low ELF value of 0.54, which dominate the DOS at $\varepsilon_{\mathrm{F}}$ \cite{LaH10_Mechanism}. Crucially, the Sc-induced structural deformation results in a volumetric density of metallic H-H interactions approximately 25\% greater than that in LaH$_{10}$ [Table S3]. This finding is in excellent quantitative agreement with the observed increase in the hydrogen-derived DOS at $\varepsilon_{\mathrm{F}}$. As a consequence, the introduction of scandium effectively optimizes the hydrogen framework, thereby enhancing the superconducting potential of LaSc$_{2}$H$_{24}$.

Beyond its indirect structural influence, scandium also directly enriches the electronic states on the Fermi surface (FS) through the hybridization of its $3d$ electrons with hydrogen atoms. As illustrated by the element-projected Fermi surface (FS) [Fig. 2(a), 2(b), and S6], the states along the $\mathrm{M-K}$ and $\mathrm{\Gamma-A}$ high-symmetry lines are clearly dominated by Sc and H$_\mathrm{II}$ atoms. Notably, the Sc-derived DOS values are remarkably elevated at $\varepsilon_\mathrm{F}$, highlighting the direct involvement of scandium in metallic behavior. Analysis of the real-space wavefunctions reveals that these hybridized states are analogous to the $\sigma$- and $\pi$-bonding states observed in MgB$_{2}$ \cite{MgB2} [Fig. 2 and S4]. These states originate from the bridging role of the H$_\mathrm{II}$ atoms between adjacent Sc atoms. Specifically, along the $\mathrm{\Gamma-A}$ lines, the $3d_\mathrm{xy}$ or $3d_\mathrm{x^{2}-y^{2}}$ components of adjacent Sc atoms achieve axial (head-to-head) overlap via the intervening H$_\mathrm{II}$ atoms, forming a Sc-H$_\mathrm{II}$-Sc $\sigma$ configuration. Conversely, along the $\mathrm{M-K}$ lines, the $3d_\mathrm{zx}$ or $3d_\mathrm{zy}$ components of adjacent Sc atoms achieve lateral (side-to-side) overlap through the intervening H$_\mathrm{II}$ atoms, resulting in a Sc-H$_\mathrm{II}$-Sc $\pi$ configuration. Consequently, by establishing Sc-H hybridized states at $\varepsilon_{\mathrm{F}}$, the introduction of scandium leads to a MgB$_{2}$-like Fermi surface, marking a clear difference from LaH$_{10}$ [Fig. S9]. This fundamental change in electronic structure underlies the different superconducting behaviors of the two systems.

\begin{figure}[htb]
    \centering
    \includegraphics[width=1.0\linewidth]{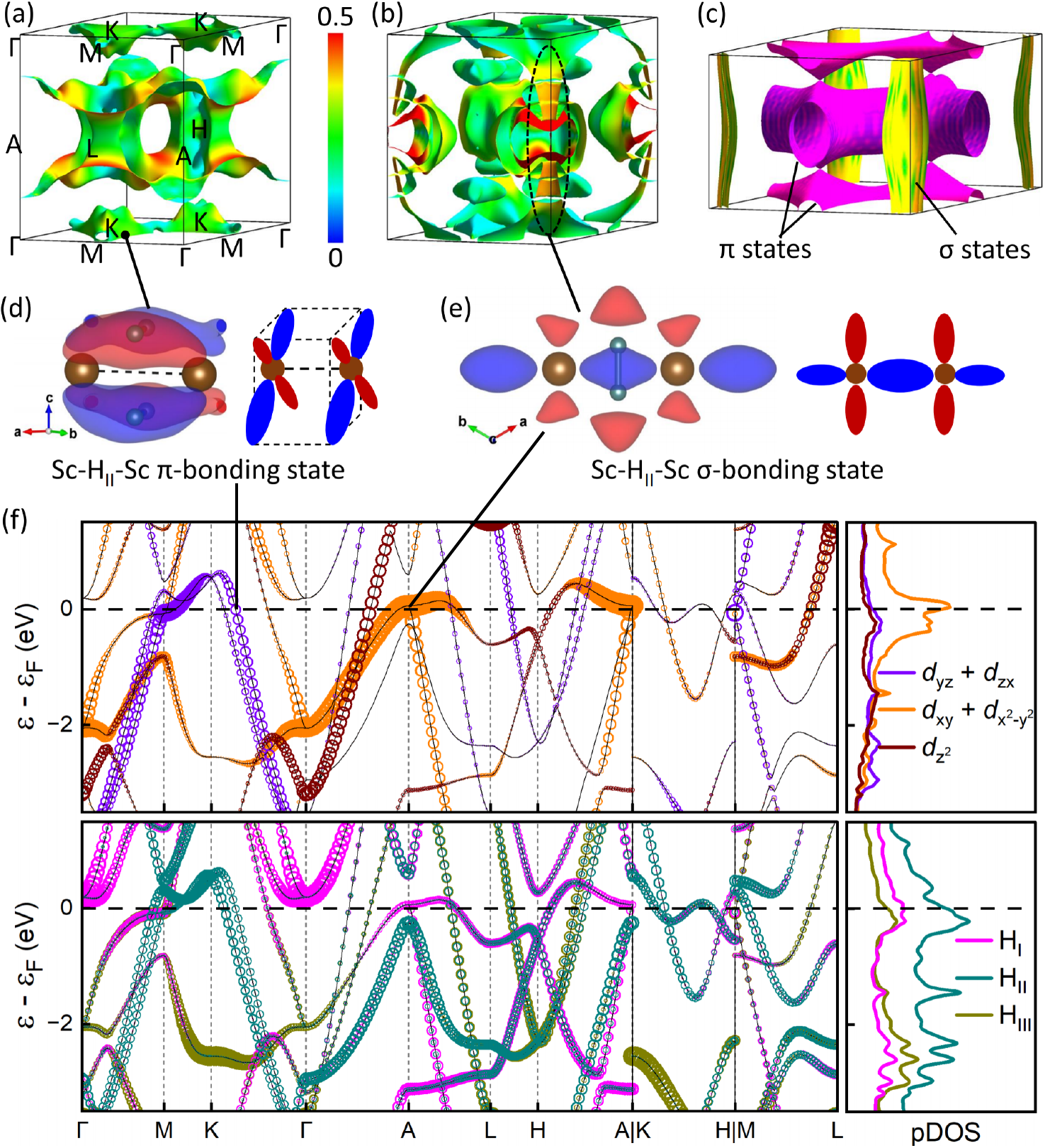}
    \caption{Element-projected Fermi surface (FS) sheets of (a) the band $n$ = 4 and (b) bands $n$ = 1, 2, 3, and 5. The Sc contribution is represented using the color scale in the range [0, 0.5]. (c) Reference FS sheets of MgB$_{2}$, showing $\sigma$ (yellow) and $\pi$ (magenta) states. (d)-(e) Orbital visualizations of the Sc-H$_\mathrm{II}$-Sc $\pi$ (lateral overlap) and $\sigma$ (axial overlap) bonding states, respectively. The real-space projections of these Sc-H states in unit cell are depicted in Supplementary Material Fig. S4 \cite{SM}. (f) Projected band structures and electron density of states (DOS) of LaSc$_{2}$H$_{24}$, annotating the typical Sc-H $\sigma$ and $\pi$ bands.}
    \label{fig:Fig. 2}
\end{figure}
To fully resolve the mechanism of room-temperature superconductivity in LaSc$_{2}$H$_{24}$, we performed a numerical solution of the anisotropic Migdal-Eliashberg (AME) equations \cite{Eliashberg, EPW1, EPW2, EPW3, EPW4}. This allowed us to calculate the distributions of $\mathbf{k}$-resolved superconducting gap ($\Delta_{n\mathbf{k}}$) and EPC constant ($\lambda_{n\mathbf{k}}$) on the Fermi surface (FS). Here, the band index $n$ and the wave vector $\mathbf{k}$ are used to distinguish electronic states. As illustrated by the energy distribution of $\Delta_{n\mathbf{k}}$ [Fig. 3(a)], LaSc$_{2}$H$_{24}$ exhibits overall isotropic single-gap superconductivity, marking a fundamental difference from the anisotropic two-gap superconductivity observed in LaH$_{10}$ [Fig. S9(a)]. This transition in superconducting behavior is clearly reflected in the distribution patterns of $\lambda_{n\mathbf{k}}$. In LaH$_{10}$, $\lambda_{n\mathbf{k}}$ is divided into two disconnected regimes: a widespread regime ($\lambda_\mathrm{1} \sim$ 2) originating from La-H hybridized states, and a strongly coupled regime ($\lambda_\mathrm{2}$ in the range of 6 to 8) derived from H-H states [Fig. S9(b)] \cite{LaH10_multigap_2020_Wang}. The variation of EPC strength in LaH$_{10}$ is primarily band-dependent, where the $\lambda_\mathrm{1}$ and $\lambda_\mathrm{2}$ regimes are distributed on FS sheets of different bands [Fig. S10(c)]. In stark contrast, $\lambda_{n\mathbf{k}}$ in LaSc$_{2}$H$_{24}$ is continuously distributed within the range of 2 to 4. Projections of $\lambda_{n\mathbf{k}}$ on each FS sheet [Fig. S7] reveal that $\lambda_{n\mathbf{k}}$ values undergo continuous variation in $\mathbf{k}$-space rather than being confined to specific bands, indicating that the EPC strength in LaSc$_{2}$H$_{24}$ is primarily $\mathbf{k}$-dependent.

The transformation in the $\lambda_{n\mathbf{k}}$ distribution originates from the modulation of the electronic structure by the introduction of scandium. On the FS of LaH$_{10}$, the strongly coupled H-H states and the widespread La-H hybridized states are present on disconnected sheets from different bands, resulting in anisotropic two-gap superconductivity. However, in LaSc$_{2}$H$_{24}$, the MgB$_{2}$-like electronic configuration ensures that Sc $3d$ electrons significantly contribute to every FS sheet [Fig. S6]. A comparison between the element-projected and $\Delta_{n\mathbf{k}}$-projected FS sheets reveals an evident inverse correlation between superconducting gap values and Sc contribution to electronic states [Fig. 2 and 3]. This demonstrates that the strongly coupled H-H states and the widespread Sc-H hybridized states are effectively merged on the Fermi surface (FS), leading to isotropic single-gap superconductivity and the $\mathbf{k}$-dependent $\lambda_{n\mathbf{k}}$. Consequently, scandium plays a critical role in the unification of the superconducting gap.

\begin{figure}[htb]
    \centering
    \includegraphics[width=1.0\linewidth]{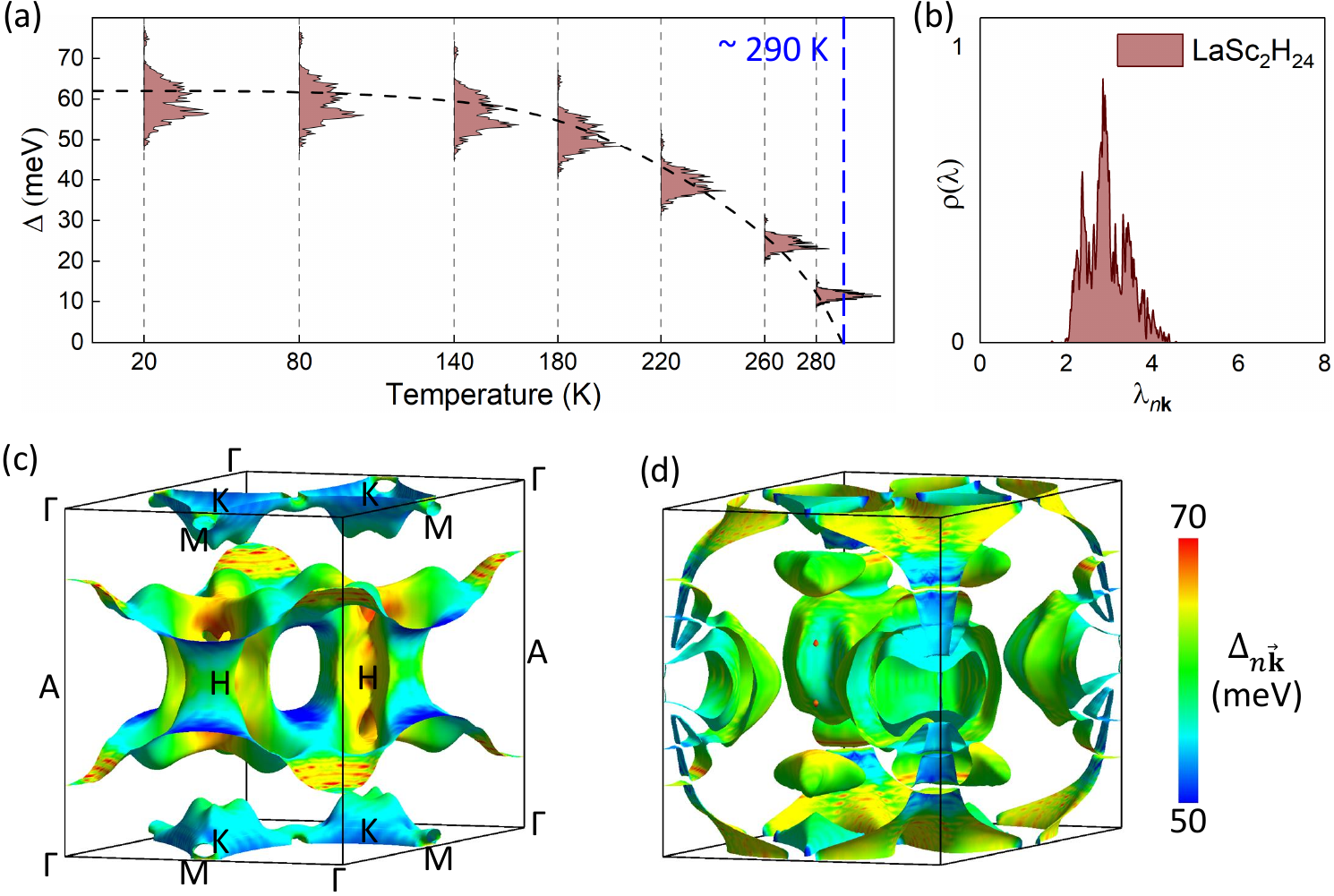}
    \caption{(a) Temperature-dependent energy distribution of superconducting gap $\Delta$ of LaSc$_{2}$H$_{24}$. The dashed curve line is guides to the eye. (b) Distribution of the EPC constant $\lambda_{n\mathbf{k}}$ of LaSc$_{2}$H$_{24}$. (c)-(d) $\mathbf{k}$-resolved superconducting gap $\Delta_{n\mathbf{k}}$ at 20 K on FS sheets of bands $n$ = 4 and $n$ = 1, 2, 3, 5, respectively. Values of $\Delta_{n\mathbf{k}}$ are represented using the color scale in the range [50, 70]. In (c) and (d), states along $\mathrm{K-H}$ lines exhibit larger $\Delta_{n\mathbf{k}}$, while along $\mathrm{\Gamma-A}$ and $\mathrm{M-K}$ lines exhibit smaller $\Delta_{n\mathbf{k}}$.}
    \label{fig:Fig. 3}
\end{figure}

To further investigate the origin of room-temperature superconductivity in LaSc$_{2}$H$_{24}$, we conducted a comprehensive study of the electronic states on the Fermi surface (FS). On one hand, the widespread hybridized states between Sc and H atoms establish the lower limit for EPC at a level comparable to LaH$_{10}$. As previously mentioned, the states along $\mathrm{\Gamma-A}$ and $\mathrm{M-K}$ lines consist of Sc-H-Sc $\sigma$- and $\pi$-bonding states. The EPC strength of these hybridized states is relatively low within LaSc$_{2}$H$_{24}$, slightly higher than that of La-H states in LaH$_{10}$ ($\lambda_{1} \sim$ 2). On the other hand, the anisotropic Sc-H hybridization enhances hydrogen metallization and introduces corresponding soft phonons, contributing significantly to the high $T_\mathrm{c}$. Specifically, the introduction of scandium enriches the H-H anti-bonding states at $\varepsilon_\mathrm{F}$. These states are localized along the $\mathrm{K-H}$ lines (indicated by red dashed lines) and exhibit significantly large $\lambda_{n\mathbf{k}}$ values [Fig. 4(c)], where the Sc-H contribution is negligibly small [Fig. S6]. Crucially, the phonon wave vector $q_\mathbf{K}$ ($\frac{1}{3}$, $\frac{1}{3}$, 0) matches the transition vector between neighboring $\mathrm{K-H}$ lines [Fig. 4(c)], and these $\mathbf{K}$-point phonon modes exhibit pronounced softening in the phonon spectrum [Fig. 4(a)]. The vibrational eigenvectors of these modes involve relative motions between different hydrogen layers along the (001) plane [Fig. 4(b)], indicating that the instability of elongated interlayer H-H bonds triggers this $\mathbf{K}$-point phonon softening. This synergy between the Fermi surface and the lattice dynamics results in a certain population of strongly coupled states with $\lambda_{n\mathbf{k}}$ values approaching 4. These two typical Sc-H and H-H states represent the dual role of Sc in superconductivity, defining both the lower and upper limits for the coupling strength. Owing to the merging of Sc-H and H-H states, the overall $\lambda$ is effectively enhanced compared to LaH$_{10}$, thereby realizing room-temperature superconductivity.

\begin{figure}[htb]
    \centering
    \includegraphics[width=1.0\linewidth]{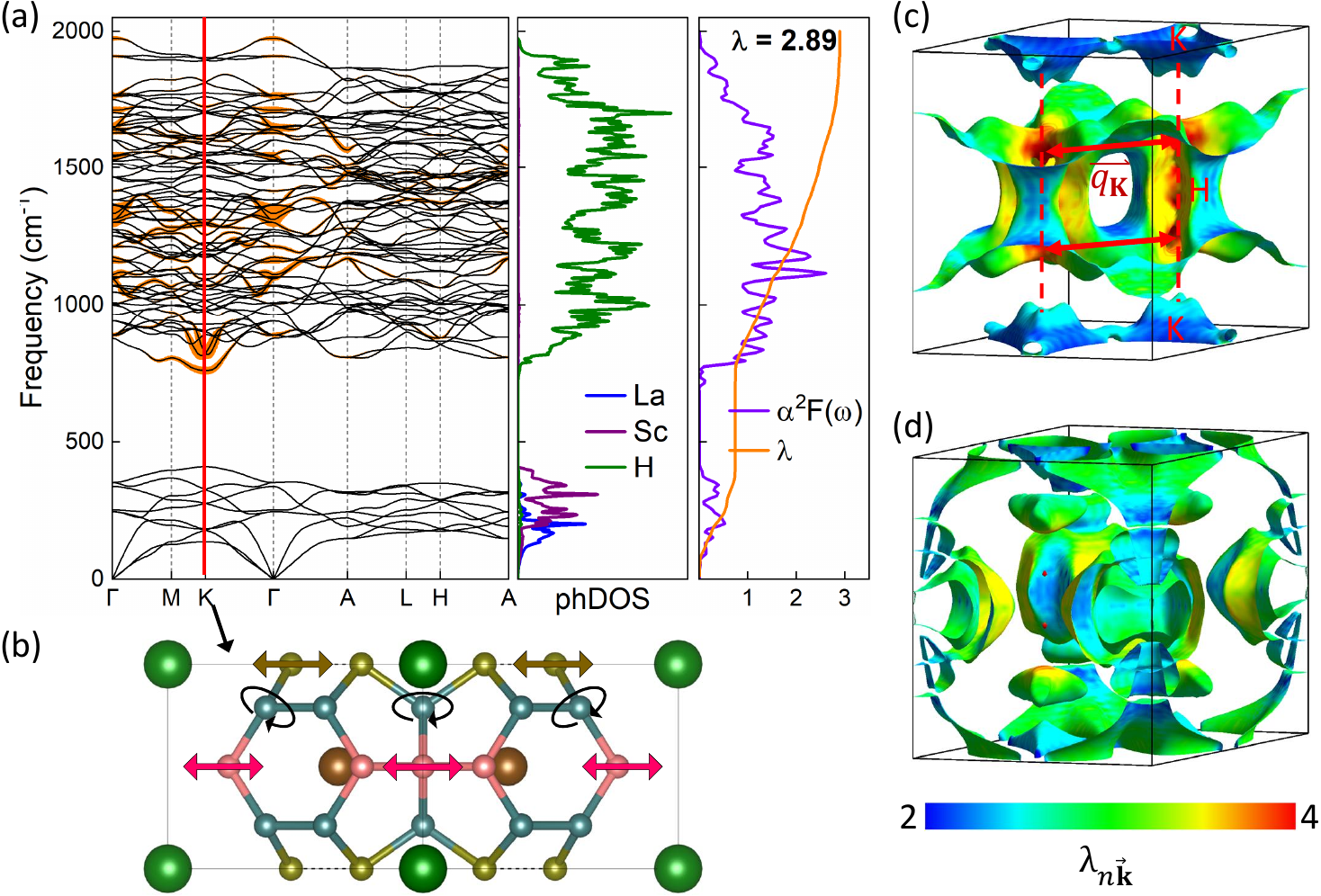}
    \caption{(a) Phonon spectrum, phonon DOS projected onto selected atoms, Eliashberg function $\alpha^{2}F(\omega)$, and integrated EPC constant $\lambda(\omega)$ of LaSc$_{2}$H$_{24}$, incorporating anharmonic effects. (b) Schematic diagram of vibrational eigenvectors of the $\mathbf{K}$-point soft phonon modes involving relative interlayer H-H motions. (c) The projected $\lambda_{n\mathbf{k}}$ on the FS sheet of band $n$ = 4, using color scale in the range [2, 4]. Herein, red regions along the $\mathrm{K-H}$ lines exhibit large $\lambda_{n\mathbf{k}}$ and translation symmetry relative to vector $q_\mathbf{K}$ ($\frac{1}{3}$, $\frac{1}{3}$, 0). (d) $\lambda_{n\mathbf{k}}$ projection on FS sheets of bands $n$ = 1, 2, 3, and 5, using color scale in the range [2, 4].}
    \label{fig:Fig. 4}
\end{figure}


In summary, we have conducted a comprehensive theoretical investigation into the mechanism of room-temperature superconductivity in the ternary hydride LaSc$_{2}$H$_{24}$ at 250 GPa. Our results underscore that the introduction of scandium drives a fundamental transition from the anisotropic two-gap superconductivity of LaH$_{10}$ to isotropic single-gap superconductivity in LaSc$_{2}$H$_{24}$. The enhancement of $T_\mathrm{c}$ originates from the dual role of the Sc $3d$ electrons. First, the Sc-induced structural deformation enhances hydrogen metallization and softens the $\mathbf{K}$-point phonon modes, significantly increasing the EPC strength of interlayer H-H states. Second, Sc introduces MgB$_{2}$-like Sc-H$_\mathrm{II}$-Sc $\sigma$- and $\pi$-bonding states on the Fermi surface (FS), providing an EPC contribution comparable to La$_{10}$. Crucially, the pronounced Sc-H hybridization ensures that the Sc-H and H-H states are effectively intermixed across the entire Fermi surface, preventing the superconducting gap separation observed in LaH$_{10}$ and establishing a continuous distribution of large EPC constants. Our findings identify this Sc-induced gap unification as the key factor for achieving superior $T_{\mathrm{c}}$, highlighting $3d$ transition metal doping as a promising strategy for engineering isotropic room-temperature superconductivity in high-pressure hydrides.

\begin{acknowledgments}

\end{acknowledgments}


\bibliography{Reference}


\end{document}